\documentclass{ws-ijmpd}

    \def\IJMP{{\it Int. J. Mod. Phys.} }

   \def\PRL{{\it Phys. Rev. Lett.} }

\begin{document}

\markboth{O. Bertolami, F. Francisco, P.J.S. Gil \& J. P\'aramos}
{Probing the flyby anomaly with the Galileo constellation}

%
%

\title{TESTING THE FLYBY ANOMALY WITH THE GNSS CONSTELLATION}

\author{ORFEU BERTOLAMI\footnote{Also at Instituto de Plasmas e Fus\~ao Nuclear, Instituto Superior T\'ecnico}}

\address{
	Departamento de F\'{\i}sica e Astronomia, Faculdade de Ci\^encias, Universidade do Porto,\\
	Rua do Campo Alegre 687, 4169-007 Porto, Portugal \\
	orfeu.bertolami@fc.up.pt}

\author{FREDERICO FRANCISCO}

\address{
	Instituto de Plasmas e Fus\~ao Nuclear, Instituto Superior T\'ecnico, Universidade T\'ecnica de Lisboa,\\
	Av.\ Rovisco Pais 1, 1049-001 Lisboa, Portugal \\
	frederico.francisco@ist.utl.pt}

\author{PAULO J. S. GIL\footnote{Also at IDMEC -- Instituto de Engenharia Mec\^anica, Instituto Superior T\'ecnico}}

\address{
	Departamento de Engenharia Mec\^anica, Instituto Superior T\'ecnico, Universidade T\'ecnica de Lisboa, \\
	Av.\ Rovisco Pais 1, 1049-001 Lisboa, Portugal \\
	p.gil@dem.ist.utl.pt}

\author{JORGE P\'ARAMOS}

\address{
	Instituto de Plasmas e Fus\~ao Nuclear, Instituto Superior T\'ecnico, Universidade T\'ecnica de Lisboa, \\
	Av.\ Rovisco Pais 1, 1049-001 Lisboa, Portugal \\
	paramos@ist.edu}

\maketitle

\begin{history}
\end{history}

\begin{abstract}
	We propose the concept of a space mission to probe the so called \emph{flyby anomaly}, an unexpected velocity change experienced by some deep-space probes using earth gravity assists. The key feature of this proposal is the use of GNSS systems to obtain an increased accuracy in the tracking of the approaching spacecraft, mainly near the perigee. Two low-cost options are also discussed to further test this anomaly: an add-on to an existing spacecraft and a dedicated mission.
\end{abstract}



\section{Introduction -- The flyby anomaly}


During the past couple of decades, a few deep-space probes that used an Earth flyby have apparently displayed an unexpected velocity change after their gravitational assist. This has become known as the \emph{flyby anomaly}.

The effect in question was detected in the residuals of the analysis performed on the Doppler and ranging data, which showed the impossibility of fitting the trajectory with a single hyperbolic arc, but allowed for a separate fit of the inward and outward paths. The additional observed velocity shift is highly localized at the perigee, where tracking through the Deep Space Network (DSN) is not available (with an approximate four hours gap). The spatial resolution of the available reconstructions, resulting form the $10~{\rm s}$ interval tracking, does not allow for an accurate characterization of the effect, so that no corresponding acceleration profile exists. Only the variation of the probes' velocity ({\it vis-\`a-vis} kinetic energy) is known.

This flyby anomaly has so far been observed in the Galileo, NEAR, Rosetta, and Cassini missions\cite{Anderson2008}. A summary of the Earth flybys observed since the 1990s is shown in Table~\ref{flyby_table}\cite{Anderson2008,Antreasian1998}. A detailed discussion of the two Galileo (1990 and 1992) and the NEAR (1998) gravity assists is available in the literature\cite{Antreasian1998}. This includes an analysis of the three earliest flybys where the anomaly was observed, with an account of the accelerations generated by different known effects, in an attempt to single out possible error sources. An estimated average acceleration associated with the flyby anomaly of the order of $10^{-4}~{\rm m/s^2}$ is measured against the Earth oblateness, other Solar System bodies, relativistic corrections, atmospheric drag, Earth albedo and infrared emissions, ocean tides, solar pressure, {\it etc}\cite{Antreasian1998}.

\begin{table}
	\tbl{Summary of orbital parameters from Earth flybys during the last couple of decades.}
	{\begin{tabular}{c c c c c c c}
		\toprule
			Mission	& Date	& $e$	& Perigee		& $v_\infty$		& $\Delta v_\infty$	& $\Delta v_\infty / v_\infty$ \\
					&		&		& $({\rm km})$	& $({\rm km/s})$ 	& $({\rm mm/s})$ 	& $(10^{-6})$ \\
		\colrule
			Galileo	& 1990	& $2.47$& $959.9$		& $8.949$			& $3.92 \pm 0.08$	& $0.438$  \\  
			Galileo	& 1992	& $3.32$& $303.1$		& $8.877$			& $-4.6 \pm 1$		& $-0.518$ \\  
			NEAR	& 1998	& $1.81$& $538.8$		& $6.851$			& $13.46 \pm 0.13$	& $1.96$   \\
			Cassini	& 1999	& $5.8$	& $1173$		& $16.01$			& $-2 \pm 1$		& $-0.125$ \\  
			Rosetta	& 2005	& $1.327$& $1954$		& $3.863$			& $1.80 \pm 0.05$	& $0.466$  \\
			MESSENGER& 2005	& -		& $2347$		& $4.056$			& $0.02 \pm 0.01$	& $0.0049$ \\
			Rosetta	& 2007	& -		& $\sim 2400$	& -					& $\sim 0$			& - \\
			Rosetta	& 2009	& -		& $2481$		& -					& $\sim 0$			& - \\
		\botrule
		\label{flyby_table}
	\end{tabular}}
\end{table}


Subsequently this discussion was extended to other possible error sources, comparing this $10^{-4}~{\rm m/s^2}$ figure with several additional unaccounted acceleration sources. These include the atmosphere, ocean tides, solid tides, spacecraft charging, magnetic moments, Earth albedo, solar wind and spin-rotation coupling. It is concluded that all of the considered effects are several orders of magnitude below the flyby anomaly\cite{Lammerzahl2006}.

A quick overview of the magnitudes of all effects discussed in the two previous paragraphs is compiled in Table \ref{error_sources_table}\cite{Antreasian1998,Lammerzahl2006}: one sees that all listed effects (except the Earth oblateness) are orders of magnitude smaller than the required value. This raises the issue of possible errors in the gravitational model of the Earth. However, attempts to solve the flyby problem by changing the related second dynamic form factor $J_2$ have yielded unreasonable solutions, and are unable to account for all flybys\cite{Antreasian1998}.

\begin{table}
	\tbl{List of orders of magnitude of possible error sources during Earth flybys.}
	{\begin{tabular}{c c}
		\toprule
			Effect				& Order of Magnitude \\
								& $({\rm m/s^2})$ \\
		\colrule
			Earth oblateness		& $10^{-2}$	\\
			Other Solar System bodies & $10^{-5}$	\\
			Relativistic effects		& $10^{-7}$	\\
			Atmospheric drag	& $10^{-7}$	\\
			Ocean and Earth tides	& $10^{-7}$	\\
			Solar pressure		& $10^{-7}$	\\
			Earth infrared		& $10^{-7}$	\\
			Spacecraft charge	& $10^{-8}$	\\
			Earth albedo		& $10^{-9}$	\\
			Solar wind			& $10^{-9}$	\\
			Magnetic moment		& $10^{-15}$	\\
		\botrule
	\end{tabular}
	\label{error_sources_table}}
\end{table}

An empirical formula to fit the flyby relative velocity change has been proposed by Anderson {\it et al.}\cite{Anderson2008} as a function of the declinations of the incoming and outgoing asymptotic velocity vectors, $\delta_i$ and $\delta_o $, respectively

\begin{equation}
	{\Delta V_\infty \over V_\infty} = K (\cos \delta_i - \cos \delta_o), \label{modelPRL}
\end{equation}
\noindent where the constant $K$ is expressed in terms of the Earth's rotation velocity $\omega_E$, its radius $R_E$ and the speed of light $c$ as 

\begin{equation}
	K = {2 \omega_e R_e \over c}.
\end{equation}

\noindent This identification is suggestive, as it evokes the general form of the outer metric due to a rotating body\cite{Ashby},

\begin{equation}
	ds^2 = \left(1 + 2{V - \Phi_0 \over c^2} \right)(c~dt)^2 - \left(1 - 2{V \over c^2} \right)(dr^2+r^2 d\Omega^2),
\end{equation}
with
\begin{equation}
	{\Phi_0 \over c^2} = {V_0 \over c^2} - {1 \over 2} \left( \omega_e R_e \over c \right)^2,
\end{equation}
where $d\Omega^2 = d\theta^2 + \sin^2\theta d\phi^2$, $V_0 $ is the value of the Newtonian potential $V(r)$ at the equator, $\omega_e$ is Earth's rotational velocity and $R_e$ is its radius. Following this reasoning, and given the strong latitude dependence of Eq. (\ref{modelPRL}), this expression appears to indicate that the Earth's rotation may be generating a much larger effect than the frame dragging predicted by General Relativity. This, however, is in contradiction with the recent measurements of this effect performed by the Gravity Probe B probe\cite{GPB}, which orbits the Earth at a height of $\sim 600~{\rm km}$, well within the onset zone of the reported flyby anomaly.


\section{Effect on GNSS systems}

In order to discuss the possible use of current and future GNSS constellations to probe this putative flyby anomaly, one should first evaluate to what extent  it can affect their individual elements. These follow approximately circular Medium Earth Orbits (MEO), at a height of about $\sim 20000 ~{\rm km}$; since the anomalous velocity change is only observed before and after flybys occurring at much smaller heights (of the order of $1000~{\rm km}$), one may empirically dismiss any effect.

One could sharpen the above argument, even though a full analysis is impossible due to the lack of spatial resolution and consequent inability to fully characterize the spatial dependence of the reported anomaly\cite{Lammerzahl2006}. Notwithstanding, one takes as relevant figure of merit the anomalous acceleration $a\sim 10^{-4}~{\rm m/s^2}$, which may be assumed constant in the absence of further data. In this case it has been shown that no constant acceleration greater than $10^{-9}~{\rm m/s^2}$ can affect the GNSS constellation, since it would have otherwise been detected\cite{Bertolami}.

Thus, one concludes that the flyby anomaly, if real, must be due to a strongly decaying force, which should drop by four orders of magnitude with a modest (about fourfold) increase in distance, from $r = R_E + h \simeq 7000~{\rm km}$ to $r \simeq 27000~{\rm km}$. As a result, one may safely assume that the GNSS constellation is fundamentally unaffected by this putative anomaly, and may be thus employed to track probes performing gravity assists at the relevant region $h \sim 1000~{\rm km}$.


\section{GNSS spacecraft tracking}

The tracking of spacecraft through GNSS systems is already commercially available ({\it e.g.} EADS-Astrium's Mosaic\cite{mosaic}, NASA PiVoT\cite{pivot}). These systems are typically used to follow satellites in low earth orbit (LEO), at altitudes below those of the GNSS satellites ($h < h_{\rm GNSS} \sim 20000~{\rm km}$), where the GNSS signal is stronger. Nevertheless, the Equator-S mission can receive front lobe signal from GPS satellites at an altitude of $61000~{\rm km}$\cite{equator}. Furthermore, it is worth exploring the possibility of using the side and back lobes of the GPS signals\cite{Mthesis,BMthesis} to establish non-line of sight tracking and avoid the shading of the Earth. Clearly, the build up of more constellations and the use of multi-GNSS receivers, able to work simultaneously with different systems, will increase the accuracy of above-MEO satellite tracking in the coming years.

The accuracy of GNSS spacecraft tracking is, understandably, better for lower orbits; however, it should be noted that during the apogee of highly elliptical orbits (HEO), the velocity is, of course, much slower than close to perigee. This allows for the construction of a good orbital solution, despite the decreased signal coverage\cite{Qiao,Moreau}. As a result, the position and velocity accuracies for different types of orbit are somewhat similar, as depicted in Table~\ref{GNSS_accuracy_table}\cite{Qiao,Moreau,Mittnacht,Kronman}.

\begin{figure}[pb]
\centering
\epsfxsize= \columnwidth
\epsffile{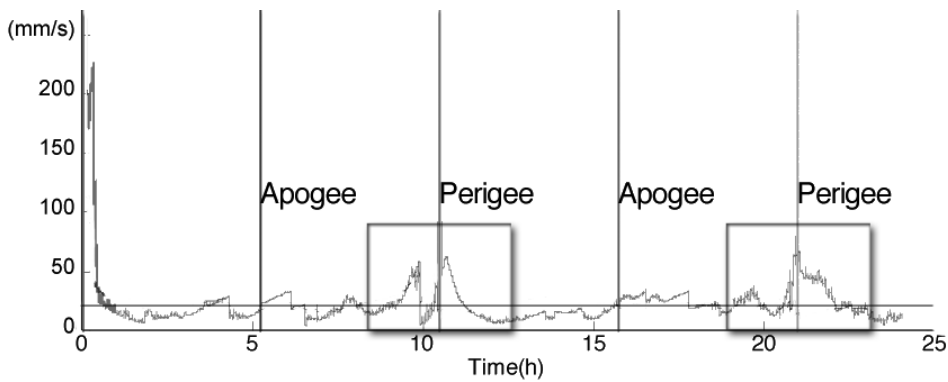}
\caption{Velocity error of multi-GNSS tracking of HEO spacecraft. Boxes (centered on perigee with 4 hour width) signal the gap in DNS coverage; the horizontal line corresponds to a $20~{\rm mm/s}$ accuracy (adapted from Ref. (12)).}
\label{velacc}
\end{figure}

Recall that there is no full characterization of the anomalies during the flyby, and these are detected from the mismatch between the expected and observed velocities after gravitational assist. As stated before, this is due to the inability of the DSN to track the spacecraft trajectories very close to the atmosphere, during a $\sim 4~{\rm h}$ gap. Regarding the possibility of using the GNSS in this region, Fig.~\ref{velacc} shows that, although the velocity error is maximum close to perigee, this peak is very localized: from a baseline of $\sim 20~{\rm mm/s}$ during the remaining orbit, it peaks briefly at $\sim 100~{\rm mm/s}$ (during the first perigee approach), and converges towards $\sim 50~{\rm mm/s}$ in the subsequent perigee passings. By plotting the aforementioned gap, one sees that accuracies of $\sim 20~{\rm mm/s}$ are attainable during approximately half of this time interval. 

For the study of the flyby anomaly, one would be interested in a high velocity accuracy, at least of the same order of magnitude as the observed $\Delta v_\infty \sim 1~{\rm mm/s}$. The currently available systems provide around $20~{\rm mm/s}$, which is clearly insufficient for such a study. However, the presented accuracies are related to real-time orbit solutions, which is unnecessary for the purpose of this study, and can undoubtedly be improved if offline processing is used, alongside other weak signal tracking strategies\cite{Moreau}. This, together with the increasing numbers of elements of the available (and upcoming) GNSS, lead us to conclude that it is indeed feasible to use the latter to test the flyby anomaly, if not with the current capability, then in the near future.

\begin{table}
	\tbl{Typical accuracies expected from GNSS satellite tracking systems for LEO, MEO, Geosynchronous Earth Orbit (GEO) and HEO.}
	{\begin{tabular}{@{} c c c c @{}} \toprule
			Orbit	& Apogee	& Position Accuracy	& Velocity Accuracy \\
					&$({\rm km})$ & $({\rm m})$			& $({\rm mm/s})$	\\
			\colrule
			LEO\cite{Mittnacht}	& $200$ to $2000$	& 10		& 10	\\
			MEO\cite{Mittnacht}	& $2000$ to GEO		& 30		& 20	\\
			GEO\cite{Mittnacht}	& $35786$			& 150		& 20	\\
			HEO\cite{Qiao}		& $> 35786$			& 100		& 20	\\ \botrule		
	\end{tabular} \label{GNSS_accuracy_table}}
\end{table}
 

\section{Options for probing the flyby anomaly}

We consider two options to test the flyby anomaly: an add-on to an existing mission on a Highly Elliptic Orbit (HEO), or a dedicated low-cost mission in either HEO or a hyperbolic trajectory.

In the first option, the choice would be to piggyback a multi-GNSS receiver in an existing space mission. Since these receivers are relatively compact and with reduced power consumption\cite{mosaic,pivot}, the host mission could be a small low-cost one. At perigee, a highly elliptical trajectory would present a comparable (although smaller) velocity and height as the reported anomalous gravitational assists, with the added benefit of allowing for repeated experiments.

One can take as an example of a suitable mission the cancelled Inner Magnetosphere Explorer (IMEX) mission of the NASA University Explorer programme, with a mass of only $ 160~{\rm kg} $ and a total budget of $ 15~{\rm M\$} $\cite{IMEX} in 2000. The IMEX probe was to be launched as a secondary payload on a Titan IV launcher, but was cancelled due to cost overrun. It would have followed a HEO, as summarized in Table \ref{tableIMEX}, which would provide a ``flyby'' velocity at perigee of about $10~{\rm km/s}$, close to the reported anomalous flybys.

The more ambitious option of a dedicated mission naturally has a number of advantages over the former, the main of which is the choice of orbit that can closely mimic a gravity assist, including an hyperbolic one. However, as discussed above, a closed orbit of sufficiently high ellipticity would provide for multiple flybys, increasing the quality of the obtained data and allowing for a better characterization of the anomaly. The HEO would also allow to ascertain if the flyby anomaly is exclusively linked to hyperbolic orbits. Also, possible error sources such as aerodynamic and thermal effects close to perigee could be more closely controlled with a dedicated mission. For instance, the spacecraft could be enclosed in a spherical radio-transparent body, so to simplify modelling and reduce directional effects. If a spin is given, any accidental anisotropies would be averaged out, yielding a much cleaner testbed for the desired experiment.

This mission would require a micro-satellite with a mass under $100~{\rm kg}$ and a budget cap similar to the IMEX mission. This upper bound is rather straightforward to argue by comparison. Firstly, no additional spending is anticipated, due to the simplified spherical design over the more complex IMEX probe. Secondly, the scientific instrumentation found in the latter would be replaced by just a multi-GNSS receiver, thus lowering the total cost. More ambitiously, an added accelerometer could provide for a cost-effective independent measure of the acceleration profile, with a modest addition to the mass budget (such as the $\sim 3~{\rm kg}$ $\mu$STAR instrument considered in the Outer Solar System and Odissey mission proposals\cite{OSS,Odissey}).

Following in the purpose of this paper, to present the feasibility of using GNSS to probe the flyby anomaly, this estimate illustrates the low cost of a dedicated mission for that purpose. Nevertheless, the actual cost could, in principle, be somewhat smaller than around $15~{\rm M \$}$, the IMEX cost estimate, not only due to the inherently simpler design and instrumentation, but also due to the ongoing trend of decreased micro-satellite costs, reflecting advances in miniaturization, lower power consumption and improved industrial processes\cite{costs}.

\begin{table}
	\tbl{Orbital parameters of IMEX's Highly Elliptical Orbit and similar hyperbolic flyby.}
	{\begin{tabular}{c c c}
		\toprule
									& IMEX	& Similar Hyperbolic Orbit \\
		\colrule
			Perigee altitude		& $349~{\rm km}$ & $349~{\rm km}$	\\
			Apogee altitude			& $35800~{\rm km}$ & --	\\
			Velocity at perigee 	& $10.1~{\rm km/s}$ &  $11~{\rm km/s}$	\\
			Eccentricity 			& $0.7248$		& $1.04$	\\
			Orbital period 			& $10.5~h$		&	--	\\			
		\botrule
	\end{tabular}
	\label{tableIMEX}}
\end{table}


\section{Conclusions}

In this work the use of the Galileo system or the GNSS to study the flyby anomaly is proposed. One finds that most of the available studies dealing with the tracking of spacecraft in real time have an insufficient velocity and position accuracy to  detect this discrepancy. However, since this real time accuracy is only one order of magnitude above the required one (in particular, $\sim 10~{\rm mm/s}$ {\it vs.} $\sim 1~{\rm mm/s}$ in velocity), it is reasonable to expect that this situation could improve in the short-term. A thorough exploitation of available resources could lead to a suitable tracking of spacecraft with greater accuracy, by abandoning real time solutions, and resorting instead to offline processing, use of side and back lobe tracking, amongst other weak signal tracking strategies. Crucially, the use of several GNSS at once should lead to an increased coverage of the different geometries.

Thus, it can be safely stated that there is no fundamental issue preventing the use of GNSS tracking to study the reported flyby anomaly. Naturally, this availability is not sufficient, as only spacecraft equipped with a (multi-)GNSS receiver would allow for such a study. In this work, we have shown that a mission of this kind could be easily deployed, either as an add-on package to an existing platform with the required highly elliptical orbit, or through a dedicated mission. While the first scenario would provide a cheap solution, we argue that a dedicated mission could be envisaged with a higher scientific payoff, while maintaining an overall low-cost approach.

Regardless of the actual origin of the flyby anomaly (unaccounted conventional effect, precision glitch or, more appealingly, new physics), we believe that our proposal offers a low-cost opportunity for displaying some of the scientific possibilities opened by the GNSS era.


\section*{Acknowledgments}

This work was partially developed in the context of the Third Colloquium {\textit Scientific and Fundamental Aspects of the Galileo Programme}, Copenhagen, August 31 - September 2, 2011. O.B. and J.P. thank the organization for the hospitality extended to them. The work of FF is sponsored by the FCT – Funda\c{c}\~{a}o para a Ci\^{e}ncia e Tecnologia (Portuguese Agency), under the grant BD 66189/2009.


\end{document}